# Students' Difficulties with Equations involving Circuit Elements

Jing Li and Chandralekha Singh

*Department of Physics and Astronomy, University of Pittsburgh, Pittsburgh, PA 15260, USA*

**Abstract.** We discuss an investigation exploring students' difficulties with equations involving resistance, capacitance and inductance. We find that introductory physics students have great difficulty understanding, e.g., how the resistance of an ohmic resistor can be written in terms of the potential difference across it and the current through it, but it does not change when the potential difference across the resistor is varied. Similar confusions arose in problems relating to capacitors and inductors. We discuss these difficulties with equations in the context of introductory physics students' performance on questions about circuit elements both in the free-response and multiple-choice formats.

## INTRODUCTION

In order to become an expert in physics, students must learn to regard an equation as a relation between physical quantities, and not merely as a plug-and-chug tool or a formula that only requires numerical substitution to obtain a solution [1-4]. They must internalize that each equation is a constraint that may relate variables and constants written in symbolic form, and that there may be many constraints relating one physical quantity to different physical quantities. They must also learn that some symbols represent universal constants, some are constant under certain conditions (e.g., the resistance of an ohmic material of a given length and a given cross sectional area at a fixed temperature), and some have a truly functional relationship (e.g., current and voltage across a resistor). Prior investigations in physics education have consistently shown that a majority of traditionally taught physics courses do not promote sense-making and conceptual reasoning; as a consequence, students often believe that problem solving in physics merely involves searching for an equation without doing a qualitative analysis and making a plan [1-4]. Here, we discuss an investigation exploring students' difficulties with equations involving circuit elements.

## METHODOLOGY

These difficulties were investigated by analyzing calculus-based introductory physics students' performance on questions about circuit elements both in the free-response and multiple-choice formats and by comparing their performance to that of physics graduate students (who were only administered the questions in the multiple-choice format due to time constraints). We also discussed the responses individually with a subset of introductory students who answered written questions. In addition to informal discussions with a subset of students who answered the written questions, we conducted formal paid interviews with six volunteers whose first midterm exam scores were close to the class average. Students who participated in the research had all received traditional lecture-based instruction on relevant content. The problems discussed here were administered in the recitations as part of quizzes in courses taught by different instructors. In addition to the written explanations and informal discussions, the analysis of the responses from the interviews yielded further information about student reasoning.

One question administered in the multiple choice format to 237 calculus-based introductory students and 42 physics graduate students was the following:

The resistance of a cylindrical ohmic resistor at a fixed temperature depends on: (I) the current; (II) the potential difference across it; (III) the cross-sectional area; (IV) the length of the resistor.

Answers: A. (I) and (II) only; B. (III) and (IV) only; C. (I), (II) and (III) only; D. (I), (II) and (IV) only; E. All of the above.

In the corresponding problem in the free-response format, various factors were listed and students had to choose all of the factors on which the resistance of an ohmic resistor at a fixed temperature depends and explain their reasoning. The free-response questions were given to 430 students from four calculus-based introductory physics courses which were different from those in which the multiple-choice questions were administered. We expected students to reason that although the resistance of a cylindrical ohmic resistor can be defined by the equation $R = V/I$, it is an intrinsic property of the resistor and is given by $R = \rho l/S$, where $V$ is the voltage, $I$ is the current, $\rho$ is the resistivity, $l$ is the length and $S$ is the cross-sectional area. We expected them to argue that the resistance does not depend on the potential difference or current. If the potential difference across the resistor is changed, the current will change correspondingly because the resistance remains fixed.

Students were asked analogous questions in both multiple-choice and open-ended formats about the capacitance of a parallel-plate capacitor and the inductance of a solenoid. Similar to the resistance question, we expected students to reason, for example, that although the capacitance is the charge on each plate per unit voltage, the ratio will remain unchanged when the voltage across the plates is changed because the charge on the plates will change correspondingly. We expected them to argue that the capacitance is an intrinsic property of a parallel plate capacitor and will depend only on the dielectric constant of the dielectric between the plates, the distance between the plates and the area of cross section of the plates.

Analogous to the resistance question, students had to choose from the following factors on which the capacitance of a parallel plate capacitor may depend: (I) the charge on the plates, (II) the potential difference across the plates, (III) the area of the plates, and (IV) the distance between the plates. The choices provided to them for the factors that determine the inductance of an inductor are (I) the current, (II) the magnetic flux through the coil and number of ideal turns of coil, (III) the cross sectional area of the coil and (IV) the number of turns per unit length. Here, the Roman numerals assigned to the factors (on which the capacitance and inductance depend) refer to their order in the multiple-choice questions and they will be referred to in the results section. We note that the resistor and capacitor questions were always administered in the same recitation but the inductor question was sometimes given in a separate recitation class after the instructor had covered the material on inductors in the course.

## RESULTS

Table 1 shows the introductory students' responses to the multiple choice questions about resistance, capacitance and inductance. It shows that 54% of the students answered the question about resistance correctly; 25% incorrectly claimed that the resistance depends on only the current and voltage; and 12% thought that the resistance depends on all of the factors given. Only 35% of the students correctly answered the question about capacitance; 29% incorrectly claimed that capacitance depends on the charge on the plates, the voltage and the distance between the plates; and 27% claimed that all of the given factors determine the capacitance. For the question about inductance, the answers were almost equally distributed across various choices suggesting students may be guessing the answer. Individual discussions with students also suggest that their knowledge about inductors was often shaky and many students admitted not understanding this topic.

**TABLE 1.** Distribution of introductory students' responses to the multiple choice questions. The numbers in bold represent the percentage of correct responses.

|             | I&II | III&IV | I,II&III | I,II&IV | ALL |
|-------------|------|--------|----------|---------|-----|
| Resistance  | 25   | **54** | 7        | 3       | 12  |
| Capacitance | 6    | **35** | 4        | 29      | 27  |
| Inductance  | 25   | **17** | 15       | 25      | 17  |

**TABLE 2.** Distribution of introductory students' responses to the free-response questions. The numbers in bold represent the percentage of correct responses.

|             | I&II | III&IV | ALL | I,III&IV | OTHER |
|-------------|------|--------|-----|----------|-------|
| Resistance  | 18   | **53** | 12  | 3        | 14    |
| Capacitance | 11   | **53** | 16  | 5        | 15    |
| Inductance  | 7    | **36** | 11  | 6        | 40    |

Table 2 shows that introductory students' response to the free-response and multiple choice questions are similar for resistance but the performance is better on the free-response version of the capacitance and inductance questions. Moreover, in the free-response questions, some students selected other combinations of the possible factors provided. Table 2 shows that for the inductor free-response question, there is again a wide variety of responses (Also, 8% chose II, III & IV, 7% choose I, II & III, and 5% choose I, II & IV.).

**TABLE 3.** Distribution of physics graduate students' responses to the multiple-choice questions. The numbers in bold represent the percentage of correct responses.

|             | I&II | III&IV | I,II&III | I,II&IV | ALL |
|-------------|------|--------|----------|---------|-----|
| Resistance  | 2    | **93** | 0        | 0       | 5   |
| Capacitance | 0    | **86** | 0        | 5       | 10  |
| Inductance  | 7    | **76** | 2        | 2       | 10  |

In order to compare and benchmark introductory students' responses with physics graduate students, we administered the same three multiple-choice questions to 42 first-year graduate students enrolled in a semester long TA training course two years in a row. Table 3 shows that while the graduate students perform significantly better than the introductory students, they have similar difficulties. Also, the inductor question is relatively difficult even for them.

## DISCUSSION

Although students performed somewhat worse on the multiple-choice questions than on the free-response ones, the results from both versions have similar trends. Written responses and individual discussions both suggest that, at least within the context of resistance, students were often more familiar with $R=V/I$ than its relation with the resistivity, length and the cross sectional area of the resistor. During individual discussions, many students were surprised that there are "two" equations for the resistance ($R=V/I$ and $R=\rho l/S$) because they felt that one should be able to plug numbers in *only* one special formula that epitomizes resistance.

Some of the students who questioned how there can be two equations for the same physical quantity (e.g., resistance) were reminded by the interviewer that the acceleration of an object can be described in terms of the net force per unit mass or the rate of change of velocity with time. In response to this comment, students often noted that they had not thought about the fact that more than one equation can be used to calculate the acceleration. Very often, however, they still continued to express their concern about the fact that a physical quantity can be calculated using two totally different equations. Discussions with a subset of students who answered the written questions and those interviewed suggest that students often believed that if one is given a formula for calculating a physical quantity, all the physics must be buried in that formula and one should be able to calculate everything about that physical quantity using that unique formula. Some of these students noted that the capacitance of a capacitor can depend on charge, voltage, as well as the area of cross section and the distance between the plates. Moreover, some of them were even able to recite the formula $C=Q/V$ and note that there was some formula that relates the capacitance to the distance between the plates and cross sectional area. These students were inconsistent in their assertion that there should be only one formula for a physical quantity. In particular, they felt that there should be only one formula for resistance but they mentioned more than one formula or relation for capacitance often without realizing that there was an inconsistency in their reasoning. The context-dependence of student reasoning is well-known in physics education research in diverse situations [1-3].

Another finding from the written tasks and discussions with a subset of students who answered the written questions and those who were formally interviewed is that students often did not think of the resistance of an ohmic resistor at a fixed temperature, the capacitance of a capacitor and the inductance of an inductor as properties of the resistor, capacitor and inductor, respectively. They often incorrectly claimed that the resistance of an ohmic resistor at a given temperature should change when the voltage or current is changed because of the definition $R=V/I$. Similarly, they claimed that the capacitance of a capacitor must depend on the potential difference across it and on the charge on the capacitor plates. As one student summarized it "if I rearrange $V=IR$ I get $R=V/I$ which means that the resistance depends on $V$ and $I$...how can it not be true?"

During individual discussions, when students were explicitly told that the resistance (of an ohmic resistor at a fixed temperature), the capacitance and the inductance are intrinsic properties of a resistor, capacitor and inductor, respectively, and were provided the relevant relationships to illustrate these points and asked to explain how they would explain relations such as $R=V/I$ or $C=Q/V$, students were confused. They were in general unable to explain, e.g., that when $V$ increases $I$ must increase proportionately in order to keep $R$ constant. Moreover, during individual discussions, in a familiar Newtonian mechanics context, the same students who correctly claimed that Newton's second law implies that increasing the net force on an object will increase the acceleration but *not* change the mass of the object, had difficulty understanding how the resistance of an ohmic resistor will not depend on the voltage and current when $V=IR$ in an abstract context.

As noted earlier, students had similar difficulties with the capacitor question and were confused about why $C=Q/V$ does not necessarily imply that $C$ depends on $Q$ and $V$. Even when the interviewer discussed both equations $C=Q/V$ and $C=\varepsilon A/d$ and asked students to interpret using both equations what $C$ should depend on, they often claimed that it should depend on all of the variables occurring on the right hand side of both equations. For example, one student noted "variables that occur in an equation affect each other". Another student noted that $C$ should depend on $Q$ and $V$ because "charge and potential difference determine whether $C$ is fully charged resulting in its ability to hold additional charge." Discussions with students (a subset of those who answered written questions and those formally interviewed) also suggest that some students believed (although such claims may be context dependent) that the capacitance depends on the energy stored in the capacitor because the voltage and the charge affect the stored energy. Even when students were told during the discussions that the capacitance is an intrinsic property of a capacitor and asked to interpret what should happen when the voltage across a capacitor is changed, it was difficult for them to exploit $C=Q/V$ to infer that $Q$ must change proportionately when $V$ changes to keep $C$ fixed. The reason could be that traditional physics courses do not typically engage students in making sense of equations and developing reasoning skills.

Similarly, many students incorrectly claimed that the inductance of an inductor depends on the current through the coil or the magnetic flux through the coil because of the equation relating the inductance to the current and the flux (although students in general admitted during individual discussions that they did not know about inductance as well as they knew about resistance and capacitance). Some students noted that the inductance depends on the current and the magnetic field because the current running through the inductor creates an induced magnetic field and the magnetic field produces the inductance. They claimed that the inductance is non-zero only when an inductor is connected in a circuit and there is a current.

During individual discussions, students were most likely to answer the inductance question using a formula perhaps because this topic was most unfamiliar to them. However, they often mixed up the definitions of magnetic flux, magnetic field, induced emf and inductance. For example, some students who used the equation for Faraday's law of electromagnetic induction confused the induced emf with the inductance of the inductor. A typical response from a student is the following: "$\mathcal{E}=-Nd\phi/dt$. The inductance depends on the number of turns and the change in magnetic flux. The flux depends on the magnetic field which depends on the current. Therefore, the inductance depends on the flux, the current and the number of turns."

## SUMMARY AND CONCLUSION

We find that, at least in some contexts, many students claimed that there should be only one equation for a physical quantity that should provide the one and only "solution" for that physical quantity. For example, they were confused when presented with two separate equations for the resistance: one relating it to the potential difference and current; and the other relating it to the resistivity, length and area of cross-section of the resistor. Similar confusions arose in problems relating to capacitors and inductors. Students often struggled with the fact that each equation is a constraint that may relate variables and constants written in symbolic form and there may be many constraints relating one physical quantity to other ones. The introductory students in general had great difficulty understanding, e.g., how the resistance of an ohmic resistor can be written in terms of the potential difference across it and the current through it, but it does not change when the voltage across the resistor is varied. It was difficult for them to understand that some symbols represent universal constants, while others are constant under certain conditions.

Instructional strategies to improve students' understanding of these issues related to interpreting equations should take into account these difficulties found in the context of circuit elements. Prior research shows that similar difficulties are prevalent across different topics [1-3]. Instruction should help students learn to reason appropriately about equations rather than viewing them as plug-and-chug tools. Physics topics (e.g., related to electrical circuit elements discussed in this paper) should not simply be taught as algorithmic exercises but rather should be used to help students develop reasoning skill. Conceptual and quantitative questions can be combined to help students do sense-making and think about the issues related to equations discussed here in more depth.

Prior research shows that knowing students' current knowledge and designing instruction to build on it is important [1-3]. While it may be easy for an instructor to understand that when the voltage increases, the current increases proportionately so that the resistance of an ohmic resistor does not change at a fixed temperature, it is challenging for students. Research suggests that it is not sufficient to tell students that $R=V/I$ does not imply that the resistance depends on the voltage or current. Students can quickly revert back to interpreting these equations incorrectly [1-3].

One strategy to help students with these concepts is to provide them with guided exploration activities within a coherent curriculum that challenge these incorrect notions and give them an opportunity to organize and extend their knowledge [5-6]. For example, within a coherent curriculum, students can be given a resistor and asked to connect it to batteries with different voltages and measure the current through it. They can be asked to calculate the ratio of $V/I$ for different cases and interpret why this ratio, which is the resistance, does not change when the voltage changes. They can perform similar explorations with a capacitor where they can increase the charge on the plates and observe how it affects the voltage across the plates and whether the ratio of the charge to the voltage is the same for different cases. They can also be asked to comment on the correctness of several statements provided to them, only one of which is correct, and discuss their reasoning with their peers and instructor.